\documentclass[a4paper,11pt]{article}
\pdfoutput=1
\usepackage{jcappub}
\usepackage{subfiles,amsmath,amssymb,bm,multirow,siunitx,graphicx,subcaption,tabularx,longtable,booktabs,hyperref,diagbox,colortbl,makecell,float,bbm,slashed,macros,cancel,mathtools}
\usepackage[normalem]{ulem}
\usepackage[T1]{fontenc}
\usepackage{xcolor}

\title{\boldmath Inverse Gertsenshtein effect as a probe of high-frequency gravitational waves}

\author[a,b]{Yutong He,}
\author[b]{Sambit K. Giri,}
\author[b]{Ramkishor Sharma,}
\author[c,d]{\\Salome Mtchedlidze}
\author[a]{and Ivelin Georgiev}

\affiliation[a]{The Oskar Klein Centre, Department of Astronomy, Stockholm University, AlbaNova, SE-10691 Stockholm, Sweden}
\affiliation[b]{Nordita, KTH Royal Institute of Technology and Stockholm University, Hannes Alfv\'ens v\"ag 12, SE-10691 Stockholm, Sweden}
\affiliation[c]{Dipartimento di Fisica e Astronomia, Università di Bologna, via Gobetti 93/2, 40122 Bologna, Italy}
\affiliation[d]{School of Natural Sciences and Medicine, Ilia State University, 3-5 Cholokashvili Ave, Tbilisi, GE-0194, Georgia
\\ \today}

\emailAdd{yutong.he@su.se}
\emailAdd{sambit.giri@su.se}
\emailAdd{ramkishor.sharma@su.se}
\emailAdd{salome.mtchedlidze@unibo.it}
\emailAdd{ivelin.georgiev@astro.su.se}

\abstract{
We apply the inverse Gertsenshtein effect, i.e., the graviton-photon conversion in the presence of a magnetic field, to constrain high-frequency gravitational waves (HFGWs).
Using existing astrophysical measurements, we compute upper limits on the GW energy densities $\OmGW$ at 16 different frequency bands.
Given the observed magnetisation of galaxy clusters with field strength $B\sim\mu\G$ correlated on $\ooo(10)\,\kpc$ scales, 
we estimate HFGW constraints in the $\ooo(10^2)\,\GHz$ regime to be $\OmGW\lesssim10^{16}$ with the temperature measurements of the Atacama Cosmology Telescope (ACT).
Similarly, we conservatively obtain $\OmGW\lesssim10^{13} (10^{11})$ in the $\ooo(10^2)\,\MHz$  ($\ooo(10)\,\GHz$) regime by assuming uniform magnetic field with strength $B\sim0.1\,\nG$ 
and saturating the excess signal over the Cosmic Microwave Background (CMB) reported by radio telescopes such as the Experiment to Detect the Global EoR Signature (EDGES), LOw Frequency ARray (LOFAR), and Murchison Widefield Array (MWA), and the balloon-borne second generation Absolute Radiometer for Cosmology, Astrophysics, and Diffuse Emission (ARCADE2)
with graviton-induced photons.
The upcoming Square Kilometer Array (SKA) can tighten these constraints by roughly 10 orders of magnitude, which will be a step closer to reaching the critical value of $\OmGW = 1$ or the Big Bang Nucleosynthesis (BBN) bound of $\OmGW\simeq1.2\times10^{-6}$.
We point to future improvement of the SKA forecast and estimate that proposed CMB measurement at the level of $\ooo(10^{0-2})\,\nK$, such as Primordial Inflation Explorer (PIXIE) and Voyage 2050, are needed to viably detect stochastic backgrounds of HFGWs.
}

\begin{document}
\maketitle
\flushbottom

\section{Introduction}
\label{sec:intro}

Gravitational waves (GWs) are a unique cosmic messenger as they propagate across vast distances unimpeded by the cosmic medium.
Astrophysical GW signals from binary mergers have been observed in the frequency ($f$) band 
from few Hz
to kHz by LIGO-Virgo-KAGRA (LVK)~\cite{LIGOScientific:2018mvr,LIGOScientific:2020ibl,LIGOScientific:2021djp},
while a stochastic GW background (SGWB) in the nHz range has recently been discovered by pulsar timing arrays (PTAs)~\cite{NANOGrav:2023gor,EPTA:2023fyk,Reardon:2023gzh,Xu:2023wog}.
{
On the largest scales, corresponding to e.g., $f\lesssim10^{-11}\,\Hz$, the cosmic microwave background (CMB) also constrains primordial SGWB in terms of the measured tensor-to-scalar ratio~\cite{BICEP2:2015nss,BICEP2:2018kqh,Planck:2018vyg,Namikawa:2019tax}.
}
With the upcoming space-based interferometers such as LISA~\cite{LISA:2017pwj},
the mHz range will also be extensively studied.
On the other hand, the high-frequency regime, broadly defined to be above kHz, is relatively less explored.
However, given that the high-frequency GWs (HFGWs) could potentially be sourced by new physics both in the early and late Universe
---
including but not limited to certain types of inflation~\cite{Kim:2004rp,Peloso:2015dsa,Domcke:2016bkh,Garcia-Bellido:2016dkw},
preheating and reheating~\cite{Figueroa:2017vfa,Kanemura:2023pnv},
beyond-Standard Model phase transitions (see Refs.~\cite{Caprini:2018mtu,Aggarwal:2020olq} and the discussions therein),
black hole (BH) superradiance~\cite{Brito:2015oca},
and light primordial black holes 
(PBH)~\cite{Dolgov:2011cq,Fujita:2014hha,Ejlli:2019bqj,Herman:2020wao,Franciolini:2022htd,Oncins:2022ydg,Gehrman:2022imk,Gehrman:2023esa}
--- it is important to explore GWs in the MHz, GHz, 
or even higher frequency ranges~\cite{Aggarwal:2020olq}.

To circumvent the lack of dedicated HFGW detectors, many indirect searches are proposed. A novel approach relies on
the inverse Gertsenshtein effect~\cite{Macedo:1983wcr, Raffelt:1987im, Cruise:2012zz, Dolgov:2012be, Ejlli:2018hke, Ejlli:2019bqj, Ejlli:2020fpt, Cembranos:2023ere},
which describes the conversion of gravitons into photons of the same frequency in the presence of a magnetic field~\cite{Gertsenshtein62, Boccaletti+70}. 
The effect is typically studied in two distinct contexts:
\begin{enumerate}
\item[\textit{(i)}] 
Astrophysical and cosmological magnetic fields:
Magnetic fields are ubiquitous as they are observed across many scales in the Universe, 
from planets and stars~\cite{Stevenson10,Schubert+11,Reiners2012} to galaxy and cluster scales~\cite{Beck:2000dc,GovFer2004}.
Correspondingly, graviton-photon conversion is studied in the presence of planetary magnetospheres~\cite{Liu:2023mll}, highly magnetised objects such as neutron stars, pulsars,
magnetars, and BHs~\cite{Dolgov:2017bpj,Feng:2022dph,
Kushwaha:2022twx,Ito:2023fcr,Kushwaha:2023poh}
,
Milky Way magnetic fields~\cite{Ramazanov:2023nxz} and 
large-scale magnetic fields originating in the early Universe, i.e., primordial (cosmological) magnetic fields (PMFs)~\cite{Pshirkov:2009sf,Dolgov:2012be,Domcke:2020yzq}.
\item[\textit{(ii)}] Laboratory settings: 
Although experiments specifically dedicated to inverse Gertsenshtein effect are yet to be conducted,
existing instruments that involve controlled magnetic fields and electromagnetic (EM) sensors, such as axion detectors, can be repurposed and their results reinterpreted as upper limits on HFGWs~\cite{Cruise:2000za, Grishchuk:2003un, Cruise:2006zt, Fuzfa:2017ana, Berlin:2021txa, Domcke:2022rgu, Domcke:2023bat, Berlin:2023grv, Vacalis:2023gdz, Schmieden:2023fzn, Bringmann:2023gba}.
\end{enumerate}
In this work, we mainly focus on the former approach but also briefly mention the latter method.
Specifically in terms of approach \textit{(i)},
we introduce the measurement of the kinematic Sunyaev-Zeldovich (kSZ) effect in galaxy clusters as a novel method for detecting HFGWs. 
This effect originates from the amplification of CMB photon energy through inverse Compton scattering by the high-energy electrons within these clusters (see Ref.~\cite{carlstrom2002cosmology} for a review).
However, the induced Gertsenshtein photons due to gravitons 
passing through galaxy cluster magnetic fields 
can also contribute to the increase of photon energy usually interpreted as the kSZ effect.
We place conservative limits on HFGWs within the frequency range of $\ooo(10^2)\,\GHz$ using the most recent observations conducted by the Atacama Cosmology Telescope (ACT)~\cite{AtacamaCosmologyTelescope:2020wtv, Amodeo:2020mmu}.

We expand the study of approach \textit{(i)} by examining the impact of the inverse Gertsenshtein effect on the radio background 
against which the cosmological 21-cm signal is measured.
This signal arises from the hyperfine splitting of ground-state neutral hydrogen atoms in the intergalactic medium (IGM) during cosmic dawn and the epoch of reionisation, corresponding to the period when the first light sources formed and reionised IGM gas (see Ref.~\cite{pritchard201221} for a review).
Previous studies have hinted at the presence of excess radio photons beyond the blackbody CMB~\cite{feng2018enhanced,Fialkov:2019vnb} to explain the 21-cm signal measurements~\cite{Bowman:2018yin}.
Despite several proposed explanations for its physical origins, including BHs~\cite{ewall2018modeling} and radio galaxies~\cite{reis2020high} at high redshift, this phenomenon remains uncertain. In this study, we investigate photons produced by the inverse Gertsenshtein effect as a potential source of this excess radio background.
Currently, several ongoing efforts, such as the Experiment to Detect the Global EoR Signature (EDGES)~\cite{Bowman:2018yin}, LOw-Frequency ARray (LOFAR)~\cite{mertens2020improved},
and Murchison Widefield Array (MWA)~\cite{trott2020deep}
are improving the measurements of the 21-cm signal during these early epochs.
And the balloon-borne second generation Absolute Radiometer for Cosmology, Astrophysics, and Diffuse Emission (ARCADE2) has also detected a radio excess beyond the 21-cm frequency.
We place upper limits on the HFGWs by conservatively assuming that they induce photons that saturate all of the reported radio excess over CMB.
In the near future, significant improvements in these measurements are expected, primarily driven by the Square Kilometre Array (SKA)~\cite{SKA}. These advancements will lead to improved constraints on the astrophysical processes governing the formation of first-generation light sources~\cite{greig201521cmmc,watkinson2022epoch}
, the properties of reionisation~\cite{Georgiev:2023yqr},
and cosmological aspects~\cite{giri2022imprints, Schneider:2023ciq}.
We forecast the potential of the SKA in indirectly constraining the HFGWs.
By comparing the upper limit constraints derived using existing measurements, forecast using SKA, and estimated using proposed future CMB surveys,
we note the general status of HFGW detection proposals and future work to improve them.

The paper is structured as follows.
In Section~\ref{sec:conversions},
we present the result of the graviton-photon conversion in the classical limit, given the context of the large-scale magnetisation of the Universe.
The detailed formulation of the effect can be found in appendix~\ref{sec:gerts}.
In Section~\ref{sec:existing_measure},
we present conservative upper bounds on HFGWs obtained from the kSZ observations with the ACT (Section~\ref{sec:kSZ}),
and from the reported excess radio background by EDGES, LOFAR, MWA, and ARCADE2 (Section~\ref{sec:excess_radio}).
In Section~\ref{sec:future}, we present forecast constraints from SKA (Section~\ref{ssec:21cm_SKA}) and future CMB surveys (Section~\ref{ssec:forecast_CMB}).
We discuss the findings in Section~\ref{sec:discussion} and conclude in Section~\ref{sec:conclusion}.

Unless otherwise stated, we set $c = \hbar = \kB = \veps_0 = 1$,
adopt the metric signature $(-+++)$,
and the gravitational coupling $\kap\equiv8\pi\GN/c^4$, with $\GN$ being Newton's constant.
We assume the standard $\Lam$CDM cosmology parameters~\cite{Planck:2018vyg}: matter parameter $\Omega_m=0.31$, baryon parameter $\Omega_b=0.049$, dark energy parameter $\Omega_\Lambda\simeq 1-\Omega_m$, Hubble constant $H_0 (100h)=67$ km/(Mpc$\cdot$s), and the blackbody CMB temperature $\TCMB\simeq2.725\,\K$.

\section{Graviton-photon conversion}
\label{sec:conversions}

We will describe the theory behind the conversion of gravitons to photons at galaxy cluster and cosmological scales in Section~\ref{ssec:galaxy_conversion} and \ref{ssec:cosmic_conversion} respectively. In Section~\ref{ssec:CMB_distortion}, we will discuss the implications of such conversions on the CMB.

\subsection{Galaxy clusters}
\label{ssec:galaxy_conversion}

We briefly discuss here the framework in which gravitons are converted into photons.
GWs with frequency $f$ traversing through a 
galaxy-cluster magnetic field
with amplitude $B$ and coherent scale (or equivalently correlation length) $l_\cor$\footnote{Note that we directly take the observationally inferred values of $l_\cor$ here.
In numerical simulations with a known magnetic field energy spectrum $E_{\rm B}(k)$ in wavenumber $k$ space, 
the correlation length $l_\cor$ is instead computed via the expression $l_\cor = \int_0^\infty\d k k^{-1} E_{\rm B}(k)\big/\int_0^\infty E_{\rm B}(k)\d k$.} can be converted into photons of the same frequency with a small but nonzero probability $\ppp_{g\map\gam}$ 
(see appendix~\ref{sec:gerts} for the detailed derivation)
\begin{equation}
\ppp_{g\map\gam}\approx5.87\times10^{-29}\Big(\frac{B}{1\,\mu\G}\Big)^2\Big(\frac{f}{100\,\GHz}\Big)^2\Big(\frac{10^{-3}\,\cm^{-3}}{\ne}\Big)^2\Big(\frac{D}{1\,\Mpc}\Big)\Big(\frac{10\,\kpc}{l_\cor}\Big),
\label{eqn:ppp_norm_astro}
\end{equation}
where $\ne$ is the electron density of the medium,
and $D$ is the total travel distance of GWs.
We assume $D>l_\cor$ and a constant field strength within the coherence scale of the magnetic field.
Equation~\eqref{eqn:ppp_norm_astro} shows that the conversion probability depends on the strength and structure of the magnetic field. 
Both strong ($B> 1\,\mu\G$~\cite{Clarke2004,Murgia:2004zn,Vogt:2005xf,Govonietal2017}) and weaker fields ($B\lesssim 0.1-1 \,\mu\G$~\cite{FuscoFemiano2001,FuscoFemiano2004}) have been observed to be correlated over scales of $\ooo(10)\,\kpc$ in 
galaxy clusters and in the dilute plasma between galaxies, known as the intracluster medium (ICM) (see Refs.~\cite{Brueggen+12,vanWeeren:2019vxy} for reviews).
These are obtained using various methods such as Faraday rotation measurements~\cite{Cooper1962,Burn1966} or the equipartition hypothesis for the observed cluster-scale, diffuse synchrotron emission~\cite{Brueggenetal2021,Missagliaetal2022}\footnote{The Faraday rotation effect refers to the change of the intrinsic polarisation plane of the polarised emission as the light passes through the magnetised medium, and is widely used for inferring the magnetic field strength and structure both on galaxy cluster scales and in the rarefied cosmic regions~\cite{Vernstrometal2019,OSullivanetal2020}. The synchrotron emission, observed in the radio waveband, traces cluster magnetic fields as well as magnetic fields extending beyond galaxy clusters~\cite{Govonietal2019,Botteonetal2019}.}.
The origin of such large-scale correlated magnetic fields is debated. 
A commonly accepted hypothesis is that they result from the amplification of weak seed fields,
which could originate either in the early Universe, referred to as PMFs, or at later epochs, e.g., during reionisation and structure formation.

The cosmological magnetohydrodynamic (MHD) simulations, which account for the amplification of weak seed magnetic fields through the combined effects of gravitational collapse and small-scale dynamos~\cite{Marinnacietal2015,Vazzaetal2018,Dominguezetal2019,Steinwandeletal2021,Mtchedlidze:2022ewp}, reproduce ICM magnetic fields with $B\sim1\,\mu\G$ strengths (see Ref.~\cite{Donnertetal2018} for a review). However, the obtained strength as well as the coherence scale of the magnetic field at the current epoch depend on the magnetogenesis scenarios,
and on the evolution of the field in the pre- and post-recombination epochs. 
For instance, in the case of PMFs, the correlation lengths of the field can be as large as $230-400 \,\kpc$ at $z=0$~\cite{Mtchedlidze:2022ewp}.
The observational constraints are also affected by the assumption of the field topology~\cite{Murgia:2004zn,Govonietal2006,NeronovVovk2010,Pshirkovetal2016}.  
Therefore, the precise calculation of the inverse Gertsenshtein effect depends on our understanding of the magnetic field structure and strength on the relevant scales.
Having this caveat in mind, we explore in our work the conversion of gravitons into photons on 
cluster scales (Section~\ref{sec:kSZ}), 
setting $B\sim0.1\,\mu\G$ over $l_\cor\sim10\,\kpc$.

\subsection{Cosmological scales}
\label{ssec:cosmic_conversion}

Beyond the scales of galaxies and clusters,
redshift dependence of the electron distribution becomes important when considering inverse Gertsenshtein effect across cosmological epochs.
The graviton-photon conversion probability modifies to~\cite{Domcke:2020yzq}
\begin{equation}
\ppp_{g\map\gam}\approx3.78\times10^{-20}\Big(\frac{B}{0.1\,\nG}\Big)^2\Big(\frac{f}{\feq}\Big)^2\Big(\frac{1\,\Mpc}{\Del l_0}\Big)\Big(\frac{\iii(z_\ini)}{6\times10^6}\Big),
\label{eqn:ppp_norm_cosmo}
\end{equation}
where $\feq = \kB T_\CMB/(2\pi\hbar)\approx56.79\,\GHz$ is the characteristic frequency of CMB assumed to be in equilibrium,
$\Del l_0 = \min(l_{\rm eq}, l_\cor^0)$ with $l_{\rm eq}\approx95\,\Mpc$ being the comoving scale of the scalar mode entering the horizon at radiation-matter equality,
and $l_\cor^0$ being the present-day coherence length of the magnetic field,
and $\iii(z_\ini)$ is an integral determined by the ionisation fraction $X_e(z)$~\cite{Kunze:2015noa}
\begin{equation}
\iii(z_\ini) = \int_{z_\obs}^{z_\ini}\d z(1 + z)^{-3/2}X_e^{-2}(z),
\label{eqn:iii}
\end{equation}
where $z_\obs$ is the redshift at which observations are made.
In line with the assumptions made in Ref.~\cite{Domcke:2020yzq},
we consider graviton-photon conversion taking place post-CMB and therefore take $z_\ini\sim1100$ in equation~\eqref{eqn:iii} (see Ref.~\cite{Dolgov:2023zqs} for a study of pre-CMB conversion).

We note that
contrary to the case discussed in Section~\ref{ssec:galaxy_conversion}, here we specifically assume the primordial origin of the observed large-scale magnetic fields. Although the current radio telescopes detect the diffuse radio emission 
only up to a redshift of $z\sim0.9$~\cite{Linderetal2014,DiGennaroetal2021,DiGennaro2023}, the analysis of the distant blazar spectra by the Fermi-LAT and High Energy Stereoscopic System (H.E.S.S.) collaborations~\cite{Aharonianetal_2023} hints at the existence of Mpc-correlated, volume-filling magnetic fields in the IGM.
This favours primordial magnetogenesis scenarios in the inflationary~\cite{TurnerWidrow1988,Ratra1992,Kanoetal2009,Emamietal2010,TomohiroMukohyama2012} or phase-transitional~\cite{Hogan1983,Quashnocketal1989,Baymetal1996,Cheng1996} epochs. 

Uncertainties, again, remain in the understanding of the evolution of PMFs across different cosmological epochs (see Ref.~\cite{Subramanian2016} for a review).
In this work, 
to 
analyse
HFGWs on cosmological scales (Section~\ref{sec:excess_radio}), 
we choose the field properties 
normalised
in equation~\eqref{eqn:ppp_norm_cosmo},
i.e., $B\sim0.1\,\nG$ and $l_\cor\sim1\,\Mpc$,
which yield considerably higher conversion probability than in individual local structures such as galaxies or clusters shown in equation~\eqref{eqn:ppp_norm_astro}.
We justify the choice of the field strength $B\sim0.1\,\nG$ by noting that although it is orders of magnitude higher than the lower bounds derived from the blazar spectra observations, i.e., $B > 10^{-14}\,\G$ in voids~\cite{Aharonianetal_2023},
it is well within the derived values of $B < 40\,\nG$~\cite{Vernstrometal2019} and $B < 4\,\nG$~\cite{OSullivanetal2020} in the IGM, and $30\leq B\leq60\,\nG $ for filaments~\cite{Vernstrom:2021hru}.
In our future work we will take into account magnetic field characteristics in different epochs predicted or modelled by MHD simulations, i.e., weighting the integral in equation~\eqref{eqn:ppp_norm_cosmo} by the corresponding field strength and coherence scales.

\subsection{Impact on the blackbody CMB radiation}
\label{ssec:CMB_distortion}

With a small but non-vanishing conversion probability $\ppp_{g\map\gam}$,
the inverse Gertsenshtein photons introduce a small distortion $\Del F_\gam$ to the CMB photon distribution which is otherwise assumed to be in equilibrium, i.e.,
\begin{equation}
\Del F_\gam(f, T) = F_\gam(f, T) - F_{\rm eq}(f, T),
\end{equation}
where $F_\gam$ is the overall photon distribution,
and $F_{\rm eq}$ describes the blackbody distribution of CMB photons
\begin{equation}
F_{\rm eq} = \frac{1}{e^{f/\feq} - 1}.
\end{equation}
We analogously define a graviton distribution $F_g(f, T)$ to be
\begin{equation}
F_g = \frac{\pi^4}{15}\Big(\frac{\feq}{f}\Big)^4\Big(\frac{\OmGW}{\Om_\gam}\Big),
\label{eqn:Fg}
\end{equation}
such that the total GW energy is obtained as
\begin{equation}
\rho_g = \int 16\pi^2f^4 F_g(f, T)\d\ln f = \rho_\crit\int\OmGW(f)\d\ln f,
\end{equation}
where $\OmGW(f)$ is the GW energy density in units of the critical density.

The distribution functions $F_{\gam, g}$ satisfy the Boltzmann equation,
\begin{equation}
\hat L F_{\gam, g} = \pm(F_g - F_\gam)\lan\Gam_{\gam g}\ran,
\label{eqn:Boltzmann_gam_g}
\end{equation}
where $\hat L = \p_t - Hf\p_f = -H(T\p_T + f\p_f)\approx-H T\p_T$ is the Liouville operator with $H$ being the Hubble parameter,
and $\p_f F_{\gam,g} = 0$ is assumed as we focus on conversions occurring at a fixed frequency.
Here $\lan\Gam_{\gam g}\ran$ is the conversion rate,
such that it integrates to the total probability along the line-of-sight (l.o.s.) $\ppp_{g\map\gam} = \int_{\los}\lan\Gam_{\gam g}\ran \d t$.
The solution of \eqref{eqn:Boltzmann_gam_g} can be obtained as~\cite{Domcke:2020yzq}
\begin{equation}
\begin{pmatrix}
F_\gam(f, T) \\
F_g(f, T)
\end{pmatrix}
=
e^{-\ppp_{g\map\gam}}
\begin{pmatrix}
\cosh\ppp_{g\map\gam} & \sinh\ppp_{g\map\gam} \\
\sinh\ppp_{g\map\gam} & \cosh\ppp_{g\map\gam}
\end{pmatrix}
\begin{pmatrix}
F_\gam(f T_\ini/T, T_\ini) \\
F_g(f T_\ini/T, T_\ini)
\end{pmatrix},
\end{equation}
which, to the leading order in $\ooo(\ppp_{g\map\gam})$, yields the expression for $\Del F_\gam$ as
\begin{equation}
\Del F_\gam(f_0, \feq) = \big(F_g(f_\ini, T_\ini) - F_{\rm eq}\big)\ppp_{g\map\gam}.
\label{eqn:dFgam}
\end{equation}
Therefore, the fractional distortion to the photon distribution at a given frequency is
\begin{equation}
\frac{\Del F_\gam}{\Feq}(f) 
= \Big[\frac{\pi^4}{15}\Big(\frac{\feq}{f}\Big)^4\Big(\frac{\OmGW}{\Om_\gam}\Big)(e^{f/\feq} - 1) - 1\Big]\,\ppp_{g\map\gam},
\label{eqn:dFgam_Feq_OmGW}
\end{equation}
where we assumed $F_\gam = F_{\rm eq}$ at the initial time.
Inverting equation~\eqref{eqn:dFgam_Feq_OmGW} yields a constraint on the GW energy density $\OmGW$ given by $\Del F_\gam/F_\gam$ as
\begin{equation}
\OmGW = \frac{15}{\pi^4}\Om_\gam\Big(\frac{\Del F_\gam}{\Feq}\ppp_{g\map\gam}^{-1} + 1\Big)(e^{f/\feq} - 1)^{-1}\Big(\frac{\feq}{f}\Big)^{-4}.
\label{eqn:OmGW_bound_invG}
\end{equation}
Note that in similar works,
sometimes the constraints are instead placed on the characteristic GW strain $\hc$,
related to $\OmGW$ via
\begin{equation}
\hc(f) = \Big(\frac{3H_0^2}{4\pi^2}\OmGW(f)f^{-2}\Big)^{1/2}.
\end{equation}
In further sections, we will 
analyse
the measurements by directly associating the distortion of the blackbody temperature to that of the photon distribution in equation~\eqref{eqn:dFgam_Feq_OmGW},
\begin{equation}
\frac{\Del F_\gam}{F_{\rm eq}} = \frac{(e^{f/f_\gam} - 1)^{-1} - (e^{f/\feq} - 1)^{-1}}{(e^{f/\feq} - 1)^{-1}} = \frac{e^{f/\feq} - e^{f/f_\gam}}{e^{f/f_\gam} - 1},
\label{eqn:dFgam_Feq}
\end{equation}
where $f_\gam=\kB T_\gam/(2\pi\hbar)$ is the characteristic frequency of a black body with temperature $T_\gam$ and an overall distribution $F_\gam$.
Explicitly in the Rayleigh-Jeans limit ($f\ll\feq\approx56.79\,\GHz$) and the high-frequency limit ($f\gg\feq\approx56.79\,\GHz$),
equation~\eqref{eqn:dFgam_Feq} takes the forms
\begin{equation}
\frac{\Del F_\gam}{\Feq}\approx
\begin{dcases}
\frac{f_\gam - \feq}{\feq} = \frac{\Del T}{T_\CMB} 
& (f\ll\feq\approx56.79\,\GHz), \\
e^{f\big(\frac{1}{\feq}-\frac{1}{f_\gam}\big)} - 
1=e^{\frac{f}{f_\gam}\frac{\Del T}{T_\CMB}}-1
& (f\gg\feq\approx56.79\,\GHz),
\end{dcases}
\label{eqn:dT/T}
\end{equation}
where $\Del T$ denotes the excess temperature on top of the CMB blackbody spectrum.
Assuming that the inverse Gertsenshtein effect is the only mechanism causing such distortion, then $\del T_g = \Del T$,
where $\del T_g$ is the temperature of graviton-induced photons.
In reality, many mechanisms can potentially contribute and therefore $\del T_g < \Del T$ (see Section~\ref{sec:kSZ} for one such example and Sections~\ref{ssec:sens_comp} and \ref{ssec:other_methods} for discussions).

\section{Existing measurements}
\label{sec:existing_measure}

We now present the constraints on the HFGWs from currently available measurements.

\subsection{Kinematic Sunyaev-Zel'dovich effect}
\label{sec:kSZ}

The kSZ effect has been conventionally used as a method to observe and study galaxy clusters since the induced CMB distortions act as markers for the underlying electron distribution in these clusters.

\begin{figure}[tbp]
\centering
\includegraphics[width = \textwidth]{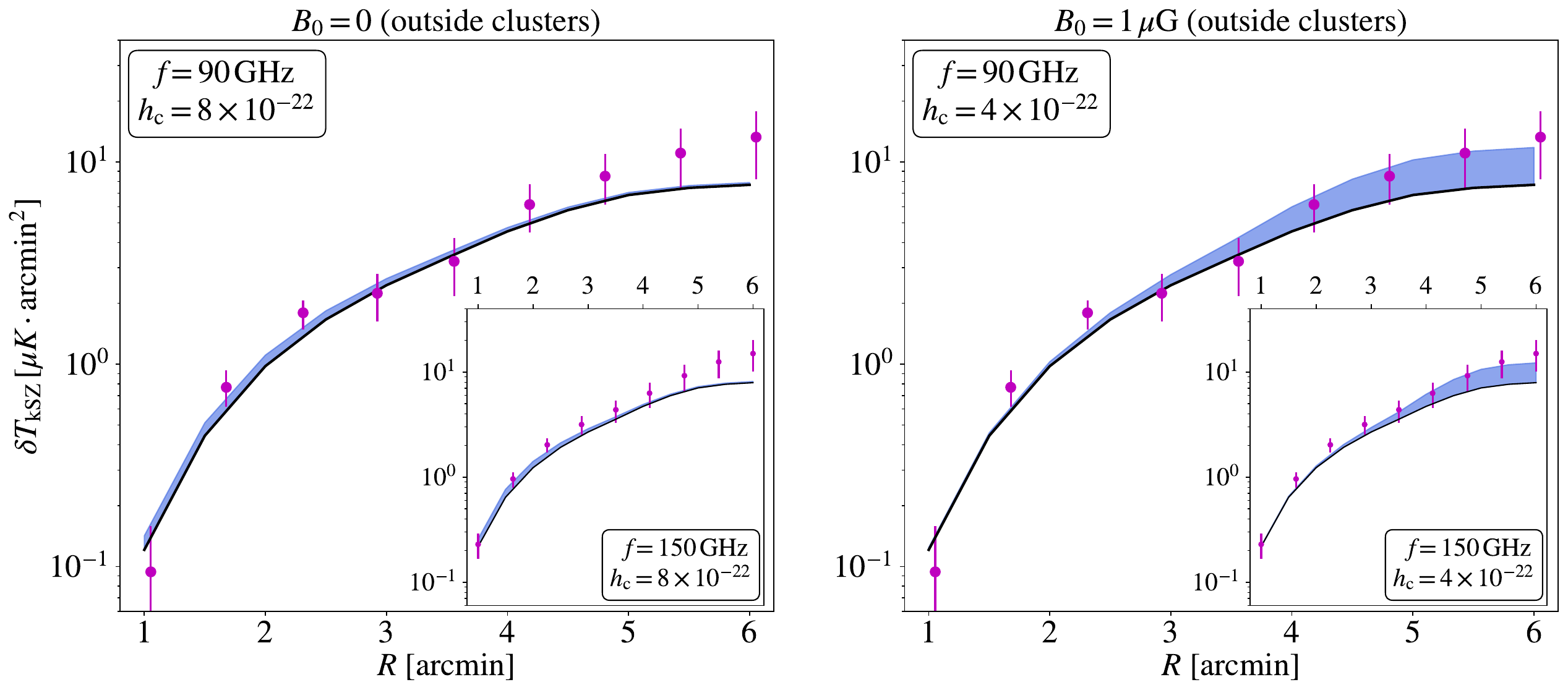}
\caption{
In both panels, ACT observations~\cite{AtacamaCosmologyTelescope:2020wtv} are indicated by the purple dots and their error bars,
and the kSZ temperature spectrum $\del\TkSZ$ from the baryon physics modelling~\cite{Schneider:2021wds} is shown in black solid lines.
The effects of graviton-induced photons are indicated by blue shaded regions, assuming the absence of magnetic fields outside clusters and a background of HFGWs with strain $\hc = 8\times10^{-22}$ (left panel) and the presence of magnetic fields with strength $B_0 = 1\,\mu\G$ and HFGWs with $\hc = 4\times10^{-22}$ (right panel).
The main panels and the insets respectively correspond to the cases of $f = 90\,\GHz$ and $f = 150\,\GHz$.}
\label{fig:T_kSZ}
\end{figure}

We provide a brief description of the kSZ modelling and refer the interested readers to Ref.~\cite{Schneider:2021wds} for more detail.
The temperature contrast observed against the CMB can be 
characterised
as~\cite{Amodeo:2020mmu}
\begin{equation}
\frac{\del\TkSZ}{\TCMB} = \frac{\sigT}{c}\int_\los e^{-\tau}\ne\vp \d l,
\label{eqn:dT_kSZ_full}
\end{equation}
where $\sigT$ is the Thomson cross-section,
$\vp$ is the peculiar velocity of the observed galaxies,
and $\tau(\theta)$ is the optical depth along the line of sight (l.o.s.).
The kSZ temperature in the above equation can be simplified in the observed redshift range of ACT, $0.4 < z < 0.7$~\cite{AtacamaCosmologyTelescope:2020wtv},
by approximating $e^{-\tau}\approx1$.
In addition, the root-mean-square (rms) peculiar velocity along the l.o.s. is $\vp^\rms/c\sim1.06\times10^{-3}$.
Therefore, equation~\eqref{eqn:dT_kSZ_full} is determined essentially by the electron density $\ne$ alone
\begin{equation}
\frac{\del\TkSZ}{\TCMB} \approx 1.06\times10^{-3}\sigT\int_\los \ne \d l,
\end{equation}
where $\ne$, in turn, depends on the gas density $\rho_\gas$. 
Modelling the baryonic physics that affects the gas distribution requires detailed hydrodynamical simulations~\cite[e.g.][]{mccarthy2016bahamas,braspenning2023flamingo}. However, we use an approximate method, called baryonic correction model (BCM), that are computationally less expensive and validated against advanced hydrodynamical simulations~\cite[]{schneider2019quantifying,giri2021emulation}. We briefly describe this model in Appendix~\ref{sec:baryonification}.
Reproducing the measurements provided by ACT requires convolving our model with the telescope filters, which are described in Ref.~\cite{AtacamaCosmologyTelescope:2020wtv}. We refer the interested readers to Refs.~\cite{Amodeo:2020mmu,Schneider:2021wds} for detailed modelling of this ACT observation.

Along with the kSZ effect, the temperature excess over the CMB can also receive a contribution from the inverse Gertsenshtein effect as HFGWs propagate through galactic and cluster magnetic fields. 
We assume that the total measured temperature contrast $\Del T = \del T_\kSZ + \del T_g$ with $\del T_g$ being the temperature of graviton-induced photons.
The effect of varying the baryon distribution inside clusters by incorporating $\del T_g$ can be seen in Figure~\ref{fig:T_kSZ}.
The black line corresponds to a BCM with $\Delta T_g=0$, aligning closely with the measurements within the 95\% credible region sourced from Ref.~\cite{Schneider:2021wds}.
The combined temperature profile $\Del T$ is tuned to stay within the error margins of ACT datasets.
We note that $\del T_g$ enhances the overall temperature profile but the exact features depend on the magnetic field properties.
Assuming that magnetic fields exist only inside the clusters and become absent in the ICM (left panel of Figure~\ref{fig:T_kSZ}), the enhancement diminishes for the corresponding scales outside individual clusters.
In this case, the strongest constraint, $\hc\geq8\times10^{-22}$, takes place at the smallest angular scale $R\approx1$, i.e., closest to the center of the cluster,
corresponding to the largest GW propagation distance.
Assuming uniform magnetic field $B_0 = 1\,\mu\G$ throughout the entire field of observation of $R\sim6\,\arcmin$, corresponding to the scale $\sim3.7\,\Mpc$ at the mean redshift $z\sim0.55$
(right panel of Figure~\ref{fig:T_kSZ}),
the temperature profile amplifies more with increasing angular scales.
In this case, GW strains are upper bounded at the largest observed scales $R\approx6$, giving $\hc\geq4\times10^{-22}$.
Therefore, detailed knowledge of magnetic fields within clusters and in the ICM is needed to determine precisely the effects of $\del T_g$,
although the results of the two assumptions shown in the left and right panels of Figure~\ref{fig:T_kSZ} differ within a factor of two.
Using equations~\eqref{eqn:OmGW_bound_invG} and~\eqref{eqn:dT/T}, and assuming the scenario of uniform magnetic fields throughout the field of observation, 
we note that the kSZ observations of ACT provide upper limits on 
the HFGW energy density $\OmGW(f)$ and strains $\hc(f)$,
i.e., $\OmGW\lesssim 3.5\times10^{15}$ and $\hc\lesssim 8\times10^{-22}$ at $f = 90\,\GHz$, and
$\OmGW\lesssim 9.9\times10^{15}$ and $\hc\lesssim 4\times10^{-22}$ at $f = 150\,\GHz$.
We find that the inverse Gertsenshtein effect is a subdominant correction on top of the more significant kSZ effect, i.e., $\del T_g \ll \del T_\kSZ$ for all observed angular scales,
even for the unrealistically large HFGW amplitudes considered here.

\subsection{Excess radio background}
\label{sec:excess_radio}

\begin{figure}[tbp]
\centering
\includegraphics[width = 0.8\textwidth]{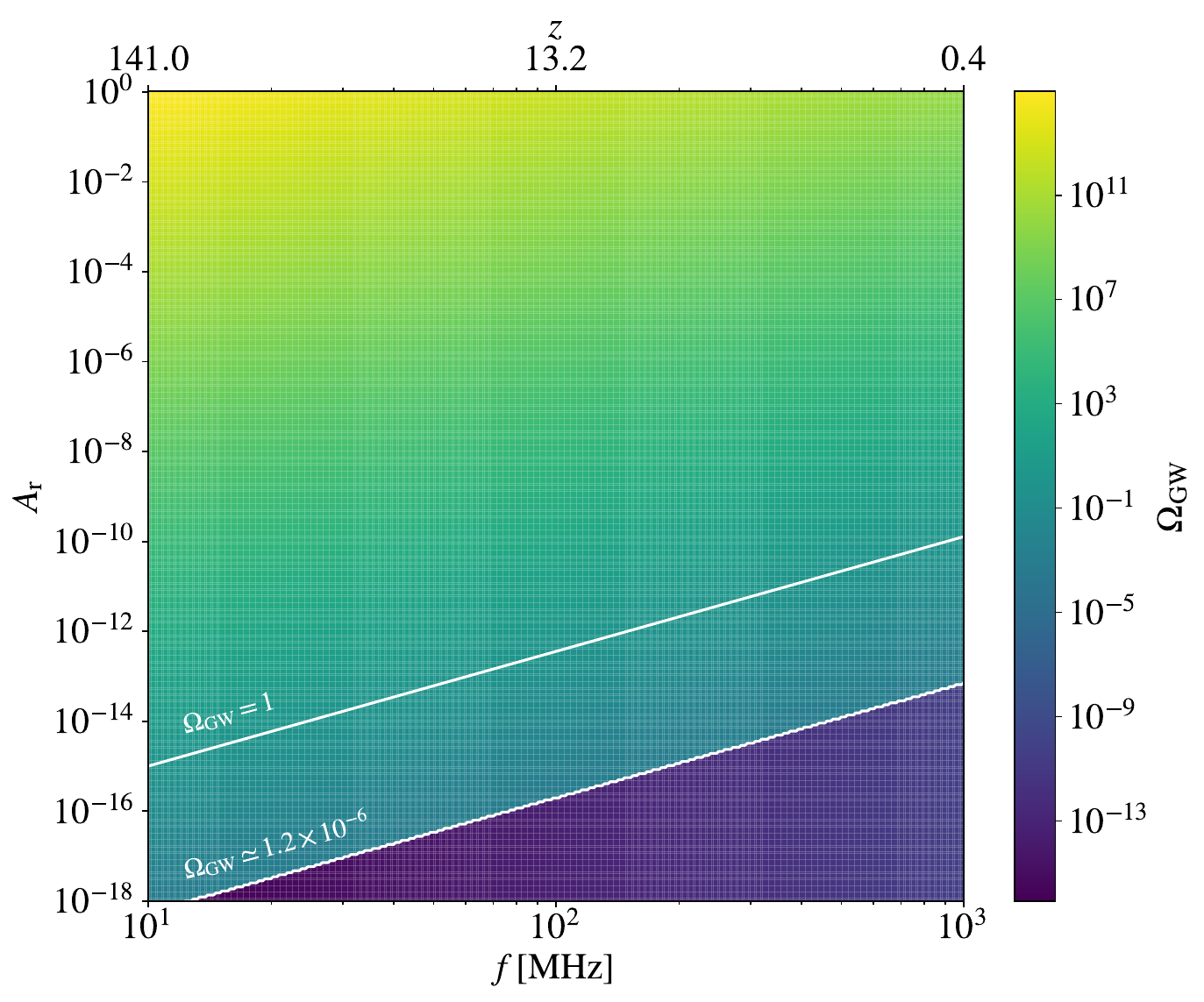}
\caption{The upper limits of GW energy density $\OmGW(f)$ imposed by a range of corresponding values of the parameter $\Ar$ in equation~\eqref{eqn:T_radio_model}, assuming cosmological magnetic fields with properties discussed in Section~\ref{ssec:cosmic_conversion}.
The two white lines respectively indicate the $\Ar$ values necessary to reach the critical value of $\OmGW = 1$ and the $\Del N_\eff$ limit of $\OmGW\simeq1.2\times10^{-6}$.
}
\label{fig:Ar_f_OmGW}
\end{figure}

Previous studies have reported an excess of radio background through various methods, including direct measurements from ARCADE2~\cite{fixsen2011arcade} and LWA1~\cite{dowell2018radio}, as well as indirect observations through the 21-cm signal during cosmic dawn.
This 21-cm signal can be observed by radio experiments
as the differential brightness temperature at position $\mathbf{x}$ and redshift $z$, which is given as~\cite{pritchard201221,SKA}
\begin{equation}
\begin{aligned}
    \delta T_{b}(\mathbf{x},z) = 27~\mathrm{mK}\left(\frac{0.15}{\Omega_{m}h^{2}}\frac{1+z}{10}\right)^{\frac{1}{2}}  \left(\frac{\Omega_{ b}h^{2}}{0.023} \right) x_{\rm HI}(\mathbf{x},z)[1+\delta_{\rm b}(\mathbf{x},z)] \left[1-\frac{T_{ \rm radio}(z)}{T_{\rm s}(\mathbf{x},z)}\right],
  \label{eq:dTb formula}
\end{aligned}
\end{equation}
where $x_\mathrm{HI}$, $\delta_\mathrm{b}$ and $T_\mathrm{s}$ are neutral hydrogen fraction, baryon overdensity and spin temperature, respectively.
This signal is observed against a radio background, which follows a blackbody spectrum of temperature $T_\mathrm{radio}$.

The previously assumed radio background to be the CMB has been challenged by measurements from ARCADE2~\cite{fixsen2011arcade} and LWA1~\cite{dowell2018radio}. These measurements indicate the presence of excess radio signal over the CMB, denoted as $T_\radio\neq T_\CMB(1 + z)$~\cite{fixsen2011arcade,dowell2018radio}. Additionally, the unconventional sky-averaged 21-cm signal evolution detected by EDGES~\cite{Bowman:2018yin} at $z\approx 17$ further supports the plausibility of this excess radio signal~\cite{Fialkov:2019vnb}. This radio background can be effectively described by the model:
\begin{equation}
T_\radio = T_\CMB(1 + z)\Big[1 + \Ar\Big(\frac{\nu_\obs}{78\,\MHz}\Big)^\bet\Big],
\label{eqn:T_radio_model}
\end{equation}
where $\Ar$ represents the frequency-independent amplitude coefficient, and $\beta\approx-2.55$ stands for the spectral index.
Note that the bounds on $\OmGW(f)$ in equation~\eqref{eqn:OmGW_bound_invG} can be rewritten in terms of the $\Ar$ parameter as
\begin{equation}
\OmGW(f) = \frac{15}{\pi^4}\Om_\gam\Big[\Ar\Big(\frac{\nu_\obs}{78\,\MHz}\Big)^\bet\ppp_{g\map\gam}^{-1} + 1\Big]\Big(\frac{\feq}{f}\Big)^{-3},
\label{eqn:OmGW_bound_invG_Ar}
\end{equation}
where the temperature distortion is given by $\Del T/T_\CMB = \Ar(\nu_\obs/78\,\MHz)^\bet$,
and we have taken the Rayleigh-Jeans limit $f\ll\feq$ applicable to the radio frequency regime.

Figure~\ref{fig:Ar_f_OmGW} demonstrates the upper bounds of $\OmGW(f)$ in equation~\eqref{eqn:OmGW_bound_invG_Ar},
given certain values of $\Ar$ and a cosmological magnetic field with properties 
normalised
in equation~\eqref{eqn:ppp_norm_cosmo}.
To give an intuition of the constraining power of $\Ar$,
we compare to two theoretical expectations,
namely the critical energy density $\OmGW = \Om_\crit = 1$,
and the more realistic $\Del N_\eff$ bound given by
\begin{equation}
\OmGW\lesssim\frac{7}{8}\Big(\frac{4}{11}\Big)^{4/3}\Om_\gam\Del N_\eff\simeq1.2\times10^{-6},
\label{eqn:OmGW_bound_Neff}
\end{equation}
where the variation of the effective number of relativistic degrees of freedom is measured to be $\Del N_\eff\lesssim0.1$~\cite{Planck:2018vyg,Cyburt:2015mya,Pagano:2015hma},
and the radiation energy density is $\Om_\gam\simeq5.4\times10^{-5}$~\cite{Planck:2018vyg}.
For example, to constrain the $\OmGW\lesssim 1(1.2\times 10^{-6})$ at $f\sim100\,\MHz$, we need to explore $\Ar\lesssim10^{-13}(10^{-16})$.

\begin{table}[tbp]
\centering
\begin{tabular}{l|r|r|r|r|c|c}
Instrument & $z$ & $f\,[\MHz]$ & $\Ar$ & $\Del T(z)\,[\rm K]$ & $\OmGW(f)$ & $\hc(f)$ \\\hline
EDGES & $17.0$ & 78 & $\lesssim 1.9$ & $\lesssim 90.49$ & $\lesssim 9.69\times10^{12}$ & $\lesssim 2.38\times10^{-20}$ \\\hline
LOFAR & $9.1$ & 141 & $\lesssim 15.9$ & $\lesssim 120.92$ & $\lesssim 1.86\times10^{13}$ & $\lesssim 1.85\times10^{-20}$ \\\hline
\multirow{6}{*}{MWA} & $8.7$ & 147 & $\lesssim 79.4$ & $\lesssim 439.71$ & $\lesssim 8.37\times10^{13}$ & $\lesssim 3.78\times10^{-20}$ \\
& $8.2$ & 155 & $\lesssim 75.9$ & $\lesssim 351.58$ & $\lesssim 6.98\times10^{13}$ & $\lesssim 3.27\times10^{-20}$ \\
& $7.8$ & 162 & $\lesssim 74.1$ & $\lesssim 296.02$ & $\lesssim 6.09\times10^{13}$ & $\lesssim 2.92\times10^{-20}$ \\
& $7.2$ & 174 & $\lesssim 25.1$ & $\lesssim 92.08$ & $\lesssim 1.72\times10^{13}$ & $\lesssim 1.45\times10^{-20}$ \\
& $6.8$ & 183 & $\lesssim 10.0$ & $\lesssim 42.69$ & $\lesssim 6.04\times10^{12}$ & $\lesssim 8.16\times10^{-21}$ \\
& $6.5$ & 190 & $\lesssim 2.5$ & $\lesssim 22.99$ & $\lesssim 1.37\times10^{12}$ & $\lesssim 3.74\times10^{-21}$\end{tabular}
\caption{Bounds of GW energy density $\OmGW(f)$, characteristic strain $\hc(f)$, and the corresponding radio temperature excess $\Del T(z)$ from existing observations of EDGES~\cite{Fialkov:2019vnb}, 
LOFAR~\cite{Mondal:2020rce}, and MWA~\cite{Ghara:2021fqo} in the $\ooo(10^2)\,\MHz$ regime.
We also show the corresponding redshift $z$ along with the $\Ar$ values constrained at 68\% confidence level.
}
\label{tab:DelT_radio_MHz}
\end{table}

We can also use the measurements from radio telescopes such as LOFAR and MWA 
to constrain $\Ar$. These telescopes are attempting to statistically measure the spatial distribution of $\delta T_b$ with the 21-cm power spectrum during the epoch of reionisation \cite[e.g.][]{mertens2020improved}. The field of view of these telescope are large enough for the signal to contain cosmological information, which has been used by previous authors to study cosmology \cite{mao2008accurately,kern2017emulating,Schneider:2023ciq}.
To provide a conservative limit, 
we assume that the entire excess radio signals were created by inverse Gertsenshtein effect.
We then use equations~\eqref{eqn:T_radio_model} and~\eqref{eqn:dT/T} to convert the values of $\Ar\neq0$ reported by authors interpreting observations from EDGES~\cite{Fialkov:2019vnb}, LOFAR~\cite{Mondal:2020rce}, and MWA~\cite{Ghara:2021fqo} to constraints on the GW energy density $\OmGW(f)$ and characteristic strain $\hc(f)$ at the corresponding frequency $f$\footnote{For the Hydrogen Epoch of Reionization Array (HERA), Ref.~\cite{HERA:2022wmy} have studied models with excess radio background. However, the constraints are provided on $T_{\rm s}/T_\radio$, which is not straightforward to convert to our parameterisation based on $\Ar$. Therefore we excluded this dataset from this work.
}.
Note that this is done assuming magnetic field properties 
normalised
in equation~\eqref{eqn:ppp_norm_cosmo}.
The resulting constraints on GWs at MHz frequencies are shown in Table~\ref{tab:DelT_radio_MHz}.
Though several orders of magnitude weaker than the $\Delta N_\mathrm{eff}$ bound, these constraints have been derived from independent measurements. Previous studies have explored several phenomena, including emission from high redshift radio galaxies~\cite{reis2020high} and superconducting cosmic strings~\cite{cyr2023cosmic}, to explain the excess radio background. If such phenomena are confirmed to contribute to this excess signal, then the constraints on GWs can be improved further. 

In Table~\ref{tab:DelT_radio_GHz}, we have listed measurements from ARCADE2~\cite{ARCADE2} that directly measure the photon background at GHz frequencies, encompassing both the CMB and GWs converted into photons. $\Del T_0$ gives the measured excess temperature over the CMB at different frequencies ($f$). This table also includes constraints on GHz frequency GWs via equation~\eqref{eqn:dT/T}, albeit being relatively weak. We anticipate improvements in these constraints with upcoming CMB experiments, which will be elaborated upon in Section~\ref{ssec:forecast_CMB}.

\begin{table}[tbp]
\centering
\begin{tabular}{l|r|r|c|c}
Instrument & $f\,[\GHz]$ & $\Del T_0\,[\K]$ & $\OmGW(f)$ & $\hc(f)$ \\\hline
\multirow{6}{*}{ARCADE2} 
& 3.2 & $\lesssim 0.0615$ & $\lesssim 2.60\times10^{11}$ & $\lesssim 9.63\times10^{-23}$ \\
& 3.4 & $\lesssim 0.0445$ & $\lesssim 2.00\times10^{11}$ & $\lesssim 7.93\times10^{-23}$ \\
& 7.9 & $\lesssim 0.0355$ & $\lesssim 3.58\times10^{11}$ & $\lesssim 4.53\times10^{-23}$ \\
& 8.3 & $\lesssim 0.0175$ & $\lesssim 1.84\times10^{11}$ & $\lesssim 3.11\times10^{-23}$ \\
& 9.7 & $\lesssim 0.0055$ & $\lesssim 6.65\times10^{10}$ & $\lesssim 1.60\times10^{-23}$ \\
& 10.5 & $\lesssim 0.0125$ & $\lesssim 1.62\times10^{11}$ & $\lesssim 2.32\times10^{-23}$
\end{tabular}
\caption{Bounds of GW energy density $\OmGW(f)$, characteristic strain $\hc(f)$, and the corresponding radio temperature excess $\Del T_0$ from ARCADE2~\cite{ARCADE2} in the $\ooo(10^0)\,\GHz$ regime.
Note that measurements of ARCADE2 at $f\in\{29.5, 31, 90\}\,\GHz$ are not shown here as they are consistent with CMB.
}
\label{tab:DelT_radio_GHz}
\end{table}

\section{Future surveys}
\label{sec:future}

We will now discuss the future surveys that are capable of improving the constraints on HFGWs. First, we present a forecast study for the upcoming radio telescope, the Square Kilometre Array (SKA). Later, we will discuss the potential of future CMB experiments.

\subsection{Forecast with the SKA}
\label{ssec:21cm_SKA}

The observation of the 21-cm signal will improve substantially with the SKA that is currently under construction at two different sites. Here we will focus on the low-frequency component built in Western Australia that will cover frequencies $f\sim 30-300$ MHz~\cite{SKA}. In this study, we will use the power spectra expected from the SKA that will quantify the spatial fluctuation strength of the 21-cm signal at different redshifts.

In order to model the power spectra, we use the analytical framework initially proposed in Ref.~\cite{schneider2021halo}. This framework has been actively improved to study the impact of cosmological structure formation on the 21-cm signal at high redshift ($z\gtrsim 6$)~\cite{giri2022imprints,Schneider:2023ciq}. 
We construct the mock observation for SKA using the open source package, \texttt{Tools21cm}~\cite{giri2020tools21cm}. 
The error in the measurement accounts for the instrumental noise that increases with increasing wavenumber ($k$). This noise was estimated assuming an observation time of 1000 hours and the latest plan of SKA antenna distribution. For a detailed description of this calculation, we refer the readers to Ref.~\cite{giri2018optimal}. We only consider the power spectra at $k\gtrsim 0.1~\mathrm{Mpc}^{-1}$ preventing the regime dominated by cosmic variance and foreground contamination~\cite{pober2014next}. 

We examine the identical mock observation illustrated in Figure 5, and the model parameters assumed during the observation listed under `Mock Value' in Table I of Ref.~\cite{Schneider:2023ciq}. These observations are constructed at 12 redshifts covering frequency $f\sim 80-200\,\MHz$.
For this mock observation, the value of $\Ar$ is zero, which corresponds to no excess radio background. To infer the constraints expected on $\Ar$, we performed an Monte Carlo Markov Chain (MCMC) analysis. We provide the full result in appendix~\ref{sec:full_mcmc_ska}. 
This analysis constrained $\Ar \lesssim 10^{-9}$ at 68\% confidence level. With magnet field properties in equation~\eqref{eqn:ppp_norm_cosmo},
i.e., $B_0\sim0.1\,\nG$ and $\Del l_0\sim1\,\Mpc$, 
this $\Ar$ value corresponds to $\OmGW\gtrsim6.0\times10^4$ at $f\sim30\,\MHz$ and $\OmGW\gtrsim4.7\times10^2$ at $f\sim200\,\MHz$. We show this constraint in Figure~\ref{fig:OmGW_f}. Compared to the current limits at these MHz frequencies (Table~\ref{tab:DelT_radio_MHz}), SKA will improve it by 
7 to 10 orders of magnitude. 

The reionisation and heating of the IGM by the first photon source or galaxies impact the spatial distribution of the signal at ionised bubble scales, which would have a scale dependent effect on the 21-cm signal~\cite{giri2019neutral,ross2021redshift,georgiev2022large}. This scale-dependent information in the SKA data will help constrain the early galaxy properties and allow constraining $\Ar$ with the amplitude of the spectrum~\cite{Mondal:2020rce}. 
However, we assume that the 21-cm background is uniformly amplified by primordial HFGWs.
Therefore, the improvement we predict for the SKA is not sensitive to the assumed properties of the first galaxies for our mock observation.
We also want to mention a caveat of our forecast study, which is the assumption that the foreground signal is assumed to be perfectly removed. 

In this study, we assumed the sensitivity of the first phase of SKA. The later stages are planned to be more sensitive that will allow the constraints on HFGWs to be even stronger. We should also note that the constraints could also be improved with longer observation time. However, processing large radio data is challenging due to various complexities, including the intricate process of data calibration (see Refs.~\cite{mevius2022numerical,gan2023assessing} and references therein).
Therefore we have considered 1000 hour observation time, which is the initial target~\cite{SKA}.
The SKA will be sensitive enough to give us measurements beyond the power spectrum, including the bispectrum~\cite[e.g.][]{giri2019position,watkinson201921} and image datasets~\cite[e.g.][]{giri2019neutral,bianco2021deep,bianco2023deep}. These measurements will have more constraining power, which we will explore in the future.

\subsection{Forecast with future CMB surveys}
\label{ssec:forecast_CMB}

\begin{table}[tbp]
\centering
\begin{tabular}{l|l|r|r|c|l|l}
Instrument  & Status & $f\,[\GHz]$ & $\Del T\,[\nK]$ & $f_\min\,[\GHz]$ & $\OmGW(f_\min)$ & $\hc(f_\min)$\\\hline
PRISTINE    & Proposed & [50, 1000] & 181 & 1000 & $1.7\times10^1$ & $2.5\times10^{-30}$ \\
PIXIE       & 
Proposed
& [10, 1000] & 13 & 1000 & $1.2\times10^2$ & $6.7\times10^{-30}$ \\
Super-PIXIE & Concept & [10, 1000] & 10 & 1000 & $9.5\times10^1$ & $5.9\times10^{-30}$ \\
Voyage 2050 & Concept & [5, 3000] & 3 & 3000 & $3.9\times10^{-13}$ & $1.3\times10^{-37}$
\end{tabular}
\caption{
Foreground-
marginalised
forecast detection capabilities of proposed 
CMB surveys PRISTINE, PIXIE/Super-PIXIE, and Voyage 2050.
The tightest HFGW bounds $\OmGW(f_\min)$ and the corresponding frequency $f_\min$ are also shown. The magnetic field properties are chosen to be in line with equation~\eqref{eqn:ppp_norm_cosmo}.}
\label{tab:forecast_constraints}
\end{table}

\begin{figure}[tbp]
\centering
\includegraphics[width = 0.8\textwidth]{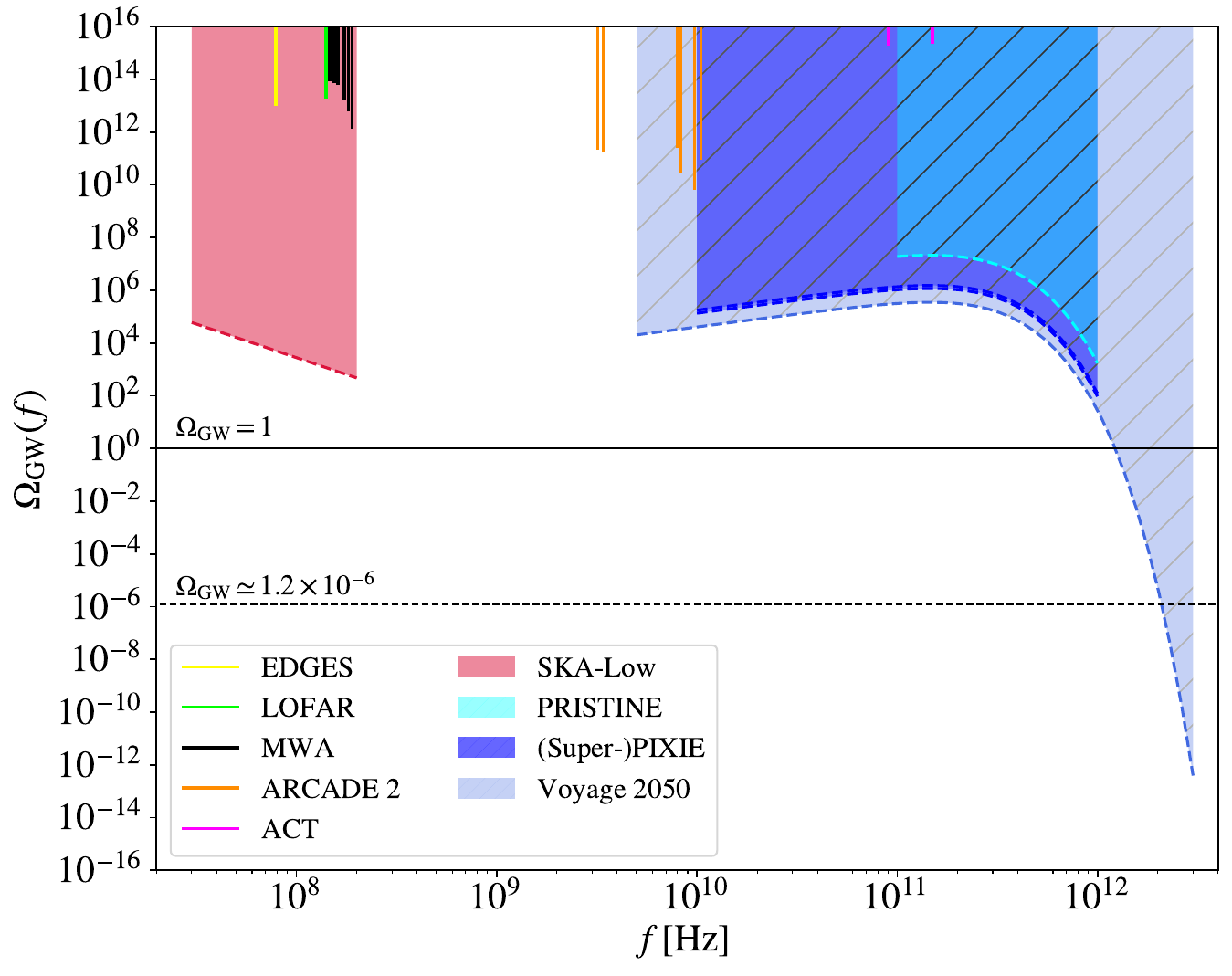}
\caption{Upper limits of HFGW energy density $\OmGW(f)$ from existing measurements using EDGES (yellow)~\cite{Bowman:2018yin,Fialkov:2019vnb}, LOFAR (green)~\cite{Mondal:2020rce}, MWA (black)~\cite{Ghara:2021fqo}, ARCADE2 (orange)~\cite{ARCADE2}, and ACT (purple)~\cite{AtacamaCosmologyTelescope:2020wtv}.
Forecast with 
the upcoming
SKA-Low (red) 
is shaded in red
.
Forecast with 
proposed and conceptual
PRISTINE (cyan)~\cite{Chluba:2019nxa}, PIXIE/Super-PIXIE (
blue
)~\cite{Kogut:2019vqh,Kogut:2011xw} and Voyage 2050 (light blue)~\cite{Basu:2019rzm} are also shown in comparison 
with tilted hatches
.
The black lines in the lower part of the figure indicate the theoretical upper bounds of GWs:
the solid line shows where $\OmGW = 1$,
and the dash-dotted line indicates the $\Del N_\eff$ bound from equation~\eqref{eqn:OmGW_bound_Neff}.
}
\label{fig:OmGW_f}
\end{figure}

In the future, a number of precision surveys of CMB and its spectral distortions could improve the constraints on HFGWs beyond the GHz regime.
Here we consider the expected capabilities of
proposed missions such as Polarized Radiation Interferometer for Spectral disTortions and INflation Exploration (PRISTINE)~\cite{Chluba:2019nxa}, 
the Primordial Inflation Explorer (PIXIE)~\cite{Kogut:2019vqh,Kogut:2011xw}
and its next-generation concept Super-PIXIE, 
as well as the scheme Voyage 2050~\cite{Basu:2019rzm},
assumed to be a few times more sensitive than Super-PIXIE.
We take the foreground-marginalized error budget for temperature measurements of these missions~\cite{Chluba:2019nxa} and compute the upper limits of HFGW energy density $\OmGW(f)$ within the corresponding detectors' frequency bands.
The anticipated temperature signal errors are shown in Table~\ref{tab:forecast_constraints} for PRISTINE (2 years), PIXIE (4 years), Super-PIXIE (8 years), and Voyage 2050~\cite{Chluba:2019nxa}.
Spectral distortions at $\ooo(10^{0-2})\,\nK$ precision, if achieved, would significantly tighten the HFGW constraints,
even reaching below the critical value $\OmGW = 1$ and the current $\Del N_\eff$ bound $\OmGW\simeq1.2\times10^{-6}$ at the THz frequency regime.
On the other hand, the $\Del N_\eff$ bound could also be tightened with these precision future surveys and a careful comparison between the expectations is needed.
The forecast constraints from SKA and future CMB surveys are shown in Figure~\ref{fig:OmGW_f} together with the estimates from Sections \ref{sec:kSZ} and \ref{sec:excess_radio}.
We show that PIXIE/super-PIXIE and, consequently, PRISTINE have the capability to significantly tighten constraints, bringing them into an intriguing range. 
Voyage 2050, which is at the conceptualisation phase, will substantially improve the constraints on HFGW probing $\OmGW\lesssim 1.2\times 10^{-6}$ in the THz regime.

\section{Discussions}
\label{sec:discussion}

In Section~\ref{ssec:sens_comp}, we compare the sensitivity of ongoing and planned experiments to constrain HFGWs. Later in Section~\ref{ssec:other_methods}, we will briefly discuss a few more alternate methods.

\subsection{Sensitivity comparison}
\label{ssec:sens_comp}

From Figure~\ref{fig:OmGW_f}, we make the following observations regarding the constraints on HFGWs from existing and projected future detectors.
\begin{itemize}
\item Among the existing instruments, 
ARCADE2 provides the tightest constraint at $\OmGW\lesssim10^{11}$,
followed by radio telescopes EDGES, LOFAR, and MWA with similar constraints at $\OmGW\lesssim10^{13}$.
The least competitive constraints come from ACT at $\OmGW\lesssim10^{16}$.
\item The constraints from ACT are the least competitive due to the low conversion probability applicable to kSZ observations of individual galaxy clusters, i.e., $\ppp\sim10^{-30}$ in equation~\eqref{eqn:ppp_norm_astro}.
This is much lower than the conversion probability applicable for global radio excess signals, i.e., $\ppp\sim10^{-20}$ in equation~\eqref{eqn:ppp_norm_cosmo}.
\item Constraints obtained in this work using existing measurements are comparable to similar work in the literature~\cite{Domcke:2020yzq,Ito:2023nkq,Liu:2023mll},
although a much tighter bound of $\OmGW < 1$ has been claimed in the X-ray frequency band~\cite{Ito:2023nkq}.
\item We found that all upcoming and future observations are significant improvements at their corresponding frequencies.
Even with an realistic amount of observing time, i.e. 1000 hours, and the conservative assumption that graviton-induced photons saturate the entire excess signals,
the SKA is expected to improve the existing constraints in the MHz frequency band by roughly 10 orders of magnitude to $\OmGW\lesssim10^{2-4}$.
\item Removing foreground contributions, proposed (PRISTINE, (Super-)PIXIE, Voyage 2050) CMB surveys can significantly tighten the $\OmGW$ upper limits, with Voyage 2050 reaching below the BBN bound in the THz regime, if the anticipated precisions are achieved. However, note that the results are highly dependent on the foreground modelling and that, as of the time of writing, these surveys remain a concept.
\item Similar to the projected CMB surveys, our conservative estimates in the $\MHz$ to $\GHz$ regimes can be further improved by combining foreground contribution from other mechanisms, such as decays of relic neutrinos~\cite{Chianese:2018luo,Dev:2023wel}, axions~\cite{Caputo:2018vmy} and other dark matter candidates~\cite{Fornengo:2011cn,Caputo:2022keo}.
We leave this to future studies.
\end{itemize}

\subsection{Alternative methods to probe HFGW}
\label{ssec:other_methods}

Though this work focused on approach \textit{(i)} introduced in Section~\ref{sec:intro}, we want to discuss the ongoing efforts to explore approach \textit{(ii)}.
Many laboratory proposals exploit the similarities between axion-photon and graviton-photon couplings to constrain HFGWs using axion detectors.
The QCD axions are pseudoscalars initially proposed as a dynamical solution to the strong charge-parity (CP) problem~\cite{Peccei:1977hh,Peccei:1977ur,Weinberg:1977ma,Wilczek:1977pj} (see Ref.~\cite{Marsh:2017hbv} for a review of axions and axion-like particles).
Numerous experimental efforts are underway to survey the parameter space of axions,
due to their desirable properties as a dark matter candidate~\cite{Preskill:1982cy,Abbott:1982af,Dine:1982ah} (see Refs.~\cite{Duffy:2009ig,Adams:2022pbo,Ferreira:2020fam} for reviews).
The coupling between axion and EM fields reads $\lll\supset g_{a\gam}aF\tilde F$,
where $a$ is the axion field and $g_{a\gam}$ the coupling strength.
This resembles the coupling between gravity and the EM sector
via the term $\lll\supset hF^2$.
The similarity implies that both axions and gravitons can convert to photons in external magnetic fields and that data from existing axion haloscopes can be reinterpreted to constrain HFGWs~\cite{Domcke:2022rgu,Domcke:2023bat}.
In general, approach \textit{(ii)} has the advantage of having full knowledge and control of the magnetic field.
The properties of large-scale magnetic fields, on the other hand, have not been precisely constrained. 
However, the future SKA surveys have the potential to substantially improve our understanding about their origin and structure on large-scales~\cite{Hollittetal2015}.
Laboratory field strength (up to $\sim10$ T) can also be much larger than cosmological fields as well, 
although the latter compensate by having a much larger effective detector volume.
Besides the two approaches of the inverse Gertsenshtein effect overall,
other methods to detect HFGWs have been proposed,
based on the couplings between gravitons and materials or mediums other than EM waves.
These include, but not limited to, 
quantum sensors to detect graviton-phonon conversion~\cite{Tobar:2023ksi,Kahn:2023mrj},
optically levitated sensors~\cite{Arvanitaki:2012cn,Aggarwal:2020umq},
microwave cavities~\cite{Caves:1979kq,Reece:1984gv,Reece:1982sc,Pegoraro:1978gv}, 
bulk acoustic wave devices~\cite{Goryachev:2014yra},
and graviton-magnon resonance detectors~\cite{Ito:2019wcb}.
Note that it is nontrivial to unify the sensitivity treatment and comparison across different detection proposals.
Therefore, we leave the quantitative comparison of the advantages and disadvantages of the proposed methods to a future work.

\section{Conclusion}
\label{sec:conclusion}

In this work we estimated the upper bounds on the stochastic background of HFGWs at 16 frequency bands in the $\ooo(10^2)\,\MHz$ and $\ooo(10^2)\,\GHz$ regimes by applying the inverse Gertsenshtein effect to large-scale magnetic fields.
The bounds are obtained conservatively by saturating with induced Gertsenshtein photons: \textit{(i)} the excess of radio background over CMB reported by EDGES, LOFAR, MWA, and ARCADE2,
and \textit{(ii)} the error margins of the kSZ observations made by ACT, assuming a fixed underlying model of baryonic physics inside galaxy clusters.
We note that these constraints are comparable in competitiveness as similar works,
and that they all lie many orders of magnitude larger than the $\Del N_\eff$ bound. 
Therefore, probing a high-frequency SGWB using inverse Gertsenshtein effect might be challenging with the existing instruments.

However, the $\Del N_\eff$ bound as a benchmark only applies to SGWB produced before BBN at $T\sim1\,\MeV$, and late-Universe mechanisms might generate GWs that reach above the bound $\OmGW\simeq1.2\times10^{-6}$ at certain frequencies.
In addition, for transient HFGW events, the theoretical bounds on a stochastic background also become invalid.
Thus, albeit the challenges, inverse Gertsenshtein effect is worth careful further studies in different contexts as a means to probe HFGWs.
It is especially so since significant improvements can be expected with future measurements. 
The upcoming SKA is expected to be much more sensitive than the current radio telescopes considered in this and other similar works.
We forecast an approximately 10 orders of magnitude tighter constraint from SKA when applied in the context of excess radio background.
Here we have only focused on the low frequency component of SKA that will observe the IGM during cosmic reionisation.
The medium frequency component of SKA will observe the 21-cm signal produced by the neutral hydrogen inside galaxies~\cite{villaescusa2018ingredients} and help fill the gap seen in Figure~\ref{fig:OmGW_f} above $f\sim200\,\MHz$. We will explore this regime in the future.
Finally, the future CMB surveys anticipated to detect spectral distortions at $\ooo(10^{0-2})\,\nK$ level could potentially reach the current $\Del N_\eff$ bound.
Finally, obtaining more realistic upper limits of HFGWs requires 
careful considerations of realistic magnetic fields, possibly using direct numerical simulations, as well as
going beyond the conservative estimations by saturating Gertsenshtein photons.
This implies a detailed understanding of other systematic and/or physical mechanisms contributing to the global foreground signals.


\acknowledgments
We thank Sreenath K. Manikandan, Florian Niedermann, and Nikhil Sarin for useful discussions.
We also thank Axel Brandenburg for a careful reading of the manuscript.
Y.H. is partially supported through the grant No.~2019-04234 from the Swedish Research Council (Vetenskapsr{\aa}det).
S.M. acknowledges financial support from the Cariplo ``BREAKTHRU'' fund (Rif: 2022-2088 CUP J33C22004310003).
Nordita is sponsored by Nordforsk.

\par\noindent\rule{\textwidth}{1pt}

\appendix
\section{Inverse Gertsenshtein formalism}
\label{sec:gerts}

In this appendix, we present some context for the inverse Gertsenshtein effect,
including a derivation of the probability equation~\eqref{eqn:ppp_norm_astro}.
See Refs.~\cite{Raffelt:1987im,Dolgov:2012be,Fujita:2020rdx,Domcke:2020yzq} for similar formulations.

\subsection*{Equations of motion}

To study the interaction between GW and EM waves,
we consider a system consisting of gravity and EM fields minimally coupled to gravity
\begin{equation}
S = S_\GR + S_\EM^{(0)} + S_\EM^{(1)},
\end{equation}
where $S_\GR$ is the Einstein-Hilbert action, and $S_\EM^{(0)}$ and $S_\EM^{(1)}$ are respectively the Maxwell and Heisenberg-Euler~\cite{Heisenberg+36} actions
\begin{align}
S_\GR & = \frac{1}{2\kap}\int\d^4x\sqrt{-g}R, \\
S_\EM^{(0)} & = -\int\d^4x\sqrt{-g}\Big(\frac{1}{4}F_\munu F^\munu - A_\mu J^\mu\Big), \\
S_\EM^{(1)} & = \frac{\alp^2}{90\me^4}\int\d^4x\sqrt{-g}\Big((F_\munu F^\munu)^2 + \frac{7}{4}(\tilde F_\munu F^\munu)^2\Big).
\end{align}
Here $F_\munu = \p_\mu A_\nu - \p_\nu A_\mu$ is the Faraday tensor,
and $\tilde F^\munu\equiv\eps^{\mu\nu\alp\bet}F_{\alp\bet}/2$ is its dual
defined using the totally antisymmetric Levi-Civita symbol $\eps^{\mu\nu\alp\bet}$,
$\alp = e^2/(4\pi)$ is the fine structure constant,
and $\me$ is the electron mass.
The Heisenberg-Euler action accounts for effective QED corrections in the low-frequency limit, $\om\ll \me\sim10^{20}\,\Hz$~\cite{Heisenberg+36,Schwinger51}.

In terms of a metric perturbation such that $g_\munu = \bar g_\munu + h_\munu$,
where $\bar g_\munu$ is the background and $h_\munu$ is a small perturbation,
we obtain the coupled Einstein-Maxwell equations of motion (EOMs) that are linear in $h_\munu$
\begin{align}
\Box h_\munu & = -2\kap T_\munu, 
\label{eqn:Einstein_full} \\
\p_\mu\Big[F^\munu - \frac{\alp^2}{45\me^4}\big(4F^2F^\munu + 7(F\cdot\tilde F)\tilde F^\munu\big)\Big] + J^\nu
& = \p_\mu(h^{\mu\bet}\bar g^{\nu\alp}F_{\bet\alp} - h^{\nu\bet}\bar g^{\mu\alp}F_{\bet\alp}),
\label{eqn:Maxwell_full}
\end{align}
where $T_\munu$ is the EM energy-momentum tensor
\begin{equation}
T_\munu = \bar g^{\rho\sig} F_{\mu\rho}F_{\nu\sig} - \frac{1}{4}\bar g_\munu F^2.
\label{eqn:EMT_EM_full}
\end{equation}
From the modified Maxwell equation~\eqref{eqn:Maxwell_full},
note that the QED corrections can be neglected if they are subdominant compared to the current.
The limit where this occurs can be obtained as follows.
We identify the EM current as 
\begin{equation}
J^\nu = -\om_\pl^2 A^\nu,
\label{eqn:J_EM}
\end{equation}
where $\om_\pl^2 = e^2 \ne/\me$ is the plasma frequency,
with $\ne$ being the electron number density.
In galaxies and clusters, the density is typically $\ne\sim10^{-3}\,\cm^{-3}$,
giving a plasma frequency of $\om_\pl\sim\ooo(10^3)\,\Hz$.

The relevance of the nonlinear QED terms in \eqref{eqn:Maxwell_full} can be estimated by noting
\begin{equation}
F^2 = 2(\BB^2 - \EE^2) = \ooo(\BB^2),
\quad
F\tilde F = 4B\cdot E \ll\ooo(\BB^2),
\end{equation}
where we have assumed $|E|\ll|B|$.
In comparison to the EM current \eqref{eqn:J_EM},
the QED corrections become important if the magnetic field satisfies
\begin{equation}
\Big(\frac{B}{B_\crit}\Big)\Big(\frac{\om}{\om_\pl}\Big)\gtrsim\frac{3}{2\alp}\sqrt{\frac{5}{2}},
\label{eqn:B_QED_limit}
\end{equation}
where $B_\crit = \me^2\sim10^{13}\,\G$ is a critical value.
In the relevant regimes we consider, 
i.e., $B\sim\mu\G$ and $\om\lesssim\ooo(10^2)\,\GHz$,
the condition \eqref{eqn:B_QED_limit} is clearly not satisfied.
Therefore, we neglect the QED corrections and
reduce the Maxwell equation~\eqref{eqn:Maxwell_full} to
\begin{equation}
(\Box - \om_\pl^2)A^\nu = - \bar g^{\mu\alp}\bar F_{\bet\alp}\p_\mu h^{\nu\bet},
\label{eqn:Maxwell_A}
\end{equation}
where we have split the Faraday tensor $F_\munu = \bar F_\munu + f_\munu$ into a quasi-static background $\bar F_\munu$ and a small perturbation $|f_\munu|\ll\bar F_\munu$ induced by GWs.
We also used the Lorenz gauge for both EM and GW quantities, 
i.e., $\p_\mu A^\mu = \p_\mu h^\munu = 0$.

\subsection*{Conversion probability}

The GW equation~\eqref{eqn:Einstein_full} in terms of its components reads
\begin{equation}
\Box
\begin{pmatrix}
h_{11} & h_{12} \\
h_{21} & h_{22} \\
\end{pmatrix}
=
-2\kap
\begin{pmatrix}
\frac{1}{2}(-B_1^2 + B_2^2 + B_3^2) & B_1B_2 \\
B_1B_2 & \frac{1}{2}(B_1^2 - B_2^2 + B_3^2) \\
\end{pmatrix},
\label{eqn:Einstein_ij_mtx}
\end{equation}
where the EM tensor \eqref{eqn:EMT_EM_full} is assumed to be dominated by the magnetic fields
\begin{equation}
T_\ij = E_iE_j + B_iB_j - \frac{1}{2}(\EE^2 + \BB^2)\del_\ij
\approx B_iB_j - \frac{1}{2}\BB^2\del_\ij.
\end{equation}
As a result of the split of the Faraday tensor,
the magnetic field 
($B = \bar B + b$) also decomposes into a homogeneous part $\bar B$ and a small induced part $|b|\ll|\bar B|$.
To bilinear order in $\bar B$ and $b$,
and let $\p_i = (0, 0, \p_l)$ in the longitudinal direction only,
then the induced magnetic fields are $(b_1, b_2, b_3) = (-\p_l A_2, \p_l A_1, 0)$ 
and equation~\eqref{eqn:Einstein_ij_mtx} becomes
\begin{equation}
\Box
\begin{pmatrix}
h_{11} & h_{12} \\
h_{21} & h_{22} \\
\end{pmatrix}
=
-2\kap
\begin{pmatrix}
(\bar B_1\p_l A_2 + \bar B_2\p_l A_1) & (\bar B_1\p_l A_1 - \bar B_2\p_l A_2) \\
(\bar B_1\p_l A_1 - \bar B_2\p_l A_2) & -(\bar B_1\p_l A_2 + \bar B_2\p_l A_1)\\
\end{pmatrix}.
\label{eqn:Einstein_ij_mtx_split}
\end{equation}
Without loss of generality,
we rotate $(B_1, B_2)\map(B, 0)$ in the transverse plane,
and denote $\lam = \{+,\times\}$ such that $h_{11} = -h_{22} = h_+, h_{12} = h_{21} = h_\times$ and $A_1 = A_\times, A_2 = A_+$.
Then the GW equation~\eqref{eqn:Einstein_ij_mtx_split} simplifies to
\begin{equation}
\Box h_\lam = -2\kap\bar B\p_l A_\lam.
\label{eqn:Einstein_fin}
\end{equation}
Similarly the Maxwell equation~\eqref{eqn:Maxwell_A} can be recast to
\begin{equation}
(\Box - \om_\pl^2)
\begin{pmatrix}
A^{(1)} \\
A^{(2)}
\end{pmatrix}
=
\begin{pmatrix}
\p_l h^{(11)} & \p_l h^{(12)} \\
\p_l h^{(21)} & \p_l h^{(22)}
\end{pmatrix}
\begin{pmatrix}
\bar B_2 \\
-\bar B_1
\end{pmatrix},
\end{equation}
which, in the same frame of $(B_1, B_2)\map(B, 0)$ as above, leads to the simplified form of
\begin{equation}
(\Box - \om_\pl^2)A_\lam = -\bar B\p_l h_\lam.
\label{eqn:Maxwell_fin}
\end{equation}
We approximate $\Box = \om^2 + \p_l^2 = (\om + i\p_l)(\om - i\p_l)\simeq 2\om(\om + i\p_l)$,
where we assumed $\p_0\simeq-i\om, -i\p_l\simeq k$ and $\om + k\simeq2\om$.
Note that we expressed the wavenumber $k$ in terms of $\omega$ using dispersion relation $k=\mu \omega$,
and used the fact that $\mu-1\ll$1 in our case to arrive at $\om + k\simeq2\om$.
We also focus on monochromatic waves such that $A_\lam(t,l) = e^{-i\om t}A_\lam(l)$ and $h_\lam(t,l) = e^{-i\om t}h_\lam(l)$.
Then the spatial solution $A_\lam(l)$ and $h_\lam(l)$ in the system of equations~\eqref{eqn:Maxwell_fin} and \eqref{eqn:Einstein_fin} can be written as
\begin{equation}
\big[\om + i\p_l - \om(1 - \mu)\big]A_\lam(l) + \frac{i}{2}\bar B h_\lam(l) =  0,\quad
(\om + i\p_l)h_\lam(l) + i\kap\bar B A_\lam(l) = 0,
\label{eqn:system_1st_order_approx}
\end{equation}
where $\mu$ is the refractive index
\begin{equation}
\mu = \sqrt{1 - (\om_\pl/\om)^2}.
\end{equation}
equations~\eqref{eqn:system_1st_order_approx} can be further summarised as
\begin{equation}
(\om + i\p_l + \mmm)\psi(l) = 0,
\label{eqn:system_soln_start}
\end{equation}
where $\psi(l) = \big(A_\lam(l), h'_\lam(l)\big)^\T$ denotes the desired solution with $h'_\lam(l) = h_\lam(l)/\sqrt{2\kap}$ (we drop the prime from here on),
and the matrix $\mmm$ reads
\begin{equation}
\mmm
=\begin{pmatrix}
\Delgamgam & -i\Delgamg \\
-i\Delgamg & 0
\end{pmatrix}
=
\begin{pmatrix}
-\om(1 - \mu) & -i\bar B\sqrt{\kap/2}\\
-i\bar B\sqrt{\kap/2} & 0
\end{pmatrix}.
\label{eqn:mtx_mmm}
\end{equation}
The relevant eigenvalues $m_{1,2}$ and the corresponding diagonal matrix are
\begin{equation}
m_{1,2} = \frac{1}{2}\Bigg[\Delgamgam\pm\sqrt{\Delgamgam^2 - 4\Delgamg^2}\Bigg], \quad
\ddd = 
\begin{pmatrix}
m_1 & 0 \\
0 & m_2
\end{pmatrix}.
\label{eqn:eigen_values_matrix}
\end{equation}
We then rotate the solution to $\psi\map\psi' = \uuu\psi$ with a unitary matrix $\uuu$,
\begin{equation}
\uuu = 
\begin{pmatrix}
\cos\theta & \sin\theta \\
-\sin\theta & \cos\theta
\end{pmatrix},\quad
\tan2\theta = \frac{2\Delgamg}{\Delgamgam},
\end{equation}
such that $\uuu^{-1} = \uuu^\T$ and $\uuu\mmm\uuu^{-1} = \ddd$.
Then in the rotated plane,
\eqref{eqn:system_soln_start} becomes
\begin{equation}
(\om + i\p_l + \ddd)\psi'(l) = 0,
\end{equation}
with its solution
\begin{equation}
\psi'(l) = e^{i(\om + \ddd)l}\psi'_\ini(l_\ini).
\end{equation}
Rotating back to the original solution space yields
\begin{equation}
\psi(l) = e^{i\om l}(\uuu^\T e^{i\ddd l}\uuu)\psi_\ini(l_\ini)
= e^{i\om l}\kkk\psi_\ini(l_\ini),
\end{equation}
where $\psi_\ini(l_\ini) = (A_{\lam,\ini}, h_{\lam,\ini})^\T$ denotes the initial conditions and the matrix $\kkk$ gives the amplitude coefficients
\begin{align}
\kkk =
\begin{pmatrix}
\kkk_{11} & \kkk_{12} \\
\kkk_{21} & \kkk_{22}
\end{pmatrix}
=
\begin{pmatrix}
[\cos^2\theta e^{im_1l} + \sin^2\theta e^{im_2l}] &
[\cos\theta\sin\theta(e^{im_1l} - e^{im_2l})] \\
[\cos\theta\sin\theta(e^{im_1l} - e^{im_2l})] &
[\sin^2\theta e^{im_1l} + \cos^2\theta e^{im_2l}]
\end{pmatrix}.
\end{align}
By starting with an initial state with only GWs and no EM waves, i.e., $\psi_{\lam,\ini} = (0, 1)^\T$,
we note that the sum of the squared coefficients in front of $h_{\lam,\ini}$ is conserved, i.e., $|\kkk_{12}|^2 + |\kkk_{22}|^2 = 1$.
Therefore, the graviton-photon conversion probability can be interpreted as the off-diagonal amplitude $|\kkk_{12}|^2$, i.e.,
\begin{equation}
P(l) = |\lan h_{\lam,\ini}|A_\lam(l)\ran|^2
= \big|\cos\theta\sin\theta\big(e^{im_1l} - e^{im_2l}\big)\big|^2
= \frac{1}{2}\kap(\bar Bl_\osc)^2\sin^2(l/l_\osc),
\label{eqn:P}
\end{equation}
where we used $\sin(\tan^{-1}\zeta) = \zeta/\sqrt{1 + \zeta^2}$,
and identified an oscillation length scale $l_\osc$ from the two eigenvalues $m_{1,2}$ in \eqref{eqn:eigen_values_matrix} such that
\begin{equation}
l_\osc^{-1} = \sqrt{\Delgamgam^2/4 + \Delgamg^2}
= \frac{1}{2}\sqrt{\om^2(1 - \mu)^2 + 2\kap\bar B^2}.
\label{eqn:l_osc}
\end{equation}
Note that \eqref{eqn:P} is suitable for a propagation distance up to the coherence scale, $l\lesssim l_\cor$, so that the uniformity of magnetic field can be assumed.
For a larger distance $D > l_\cor$,
the total conversion probability can be approximated by averaging out the sinusoidal part in equation~\eqref{eqn:P} as $1/2$ to yield
\begin{equation}
\ppp_{g\map\gam}(D)
\simeq\frac{1}{4}\kap(\bar B l_\osc)^2\Big(\frac{D}{l_\cor}\Big),
\label{eqn:ppp}
\end{equation}
which leads to the expression quoted as equation~\eqref{eqn:ppp_norm_astro}\footnote{Note that computing the numerical values requires restoring the correct dimensions in several expressions, e.g., $\om_\pl^2\map\om_\pl^2/\veps_0$ from equation~\eqref{eqn:J_EM} onward,
$\kap\map\kap/\mu_0$ from equation~\eqref{eqn:P} onward,
and $\om^2\map\om^2/c^2$ in equation~\eqref{eqn:l_osc}.},
\begin{equation}
\ppp_{g\map\gam}\approx5.87\times10^{-29}\Big(\frac{B}{1\,\mu\G}\Big)^2\Big(\frac{\om}{100\,\GHz}\Big)^2\Big(\frac{10^{-3}\,\cm^{-3}}{\ne}\Big)^2\Big(\frac{D}{1\,\Mpc}\Big)\Big(\frac{10\,\kpc}{l_\cor}\Big).
\end{equation}

\section{Modelling baryonic physics inside galaxy clusters}
\label{sec:baryonification}

We briefly elaborate on the BCM that yields the gas profile $\rho_\gas$ and electron numder density $\ne$ used for the kSZ effect calculations in Section~\ref{sec:kSZ}.
We employed a baryon model, initially presented in Ref.~\cite{schneider2015new}, and further refined in subsequent works~\cite{schneider2019quantifying,giri2021emulation}, to capture and characterise the processes inside clusters.
The matter inside the galaxy clusters consists of collisionless dark matter (clm), gas and the central galaxy matter (cga). In BCM, the total matter ($\rho_{\rm dmb}$) is given as,
\begin{equation}
\rho_{\rm dmb}(r) = \rho_{\rm clm}(r) + \rho_\gas(r) + \rho_{\rm cga}(r),
\end{equation}
where the gas profile is given by
\begin{equation}
\rho_\gas\propto\frac{\Om_\b/\Om_\mathrm{m} - f_\star/M_\vir}{\big[1 + 10(r/r_\vir)\big]^{\bet(M_\vir)}\big[1 + r/(\theta_\ej r_\vir)\big]^{[\del - \bet(M_\vir)]/\gam}},
\label{eqn:rho_gas}
\end{equation}
with $\bet$ being a mass-dependent slope,
$f_\cga$ the stellar-to-halo fractions of the central galaxy,
and $f_\star$ the total stellar content. These quantities are parameterised as,
\begin{equation}
\bet(M_\vir) = \frac{3(M_\vir/\Mc)^\mu}{1 + (M_\vir/\Mc)^\mu},\quad
f_i(M_\vir) = \frac{M_i}{M_\vir} = 0.055\Big(\frac{M_S}{M_\vir}\Big)^{\eta_i},
\end{equation}
where $i\in\{\cga,\star\}$,
$M_S = 2\times10^{11}M_\odot/h$ and the power indices are $\eta_\star = \eta$ and $\eta_\cga = \eta + \eta_\del$.
Here $\Om_\b/\Om_\mat$ is the baryon fraction,
$r_\vir$ and $M_\vir$ are the virial radius and mass.
Note that in the large-cluster limit,
equation~\eqref{eqn:rho_gas} approaches the truncated Navarro-Frenk-White (NFW) profile,
i.e., $\lim_{M_\vir\gg\Mc}\bet = 3$.
In essence, the electron number density depends on a 7-parameter model with five gas parameters $\pmb{\theta}_\mathrm{gas} \equiv (\log\Mc, \mu, \theta_\ej, \gam, \del)$, and two stellar parameters $\pmb{\theta}_\mathrm{star} \equiv (\eta, \eta_\del)$. In this study, we fixed the model with $\pmb{\theta}_\mathrm{gas}=(13.4,0.3,4,2,7)$ and $\pmb{\theta}_\mathrm{star}=(0.32,0.28)$, which is within the 95\% credible region of the constraints in Ref.~\cite{Schneider:2021wds}.

\section{Full posterior distribution of the SKA forecast study}
\label{sec:full_mcmc_ska}

\begin{figure}[tbp]
\centering
\includegraphics[width = 0.8\textwidth]{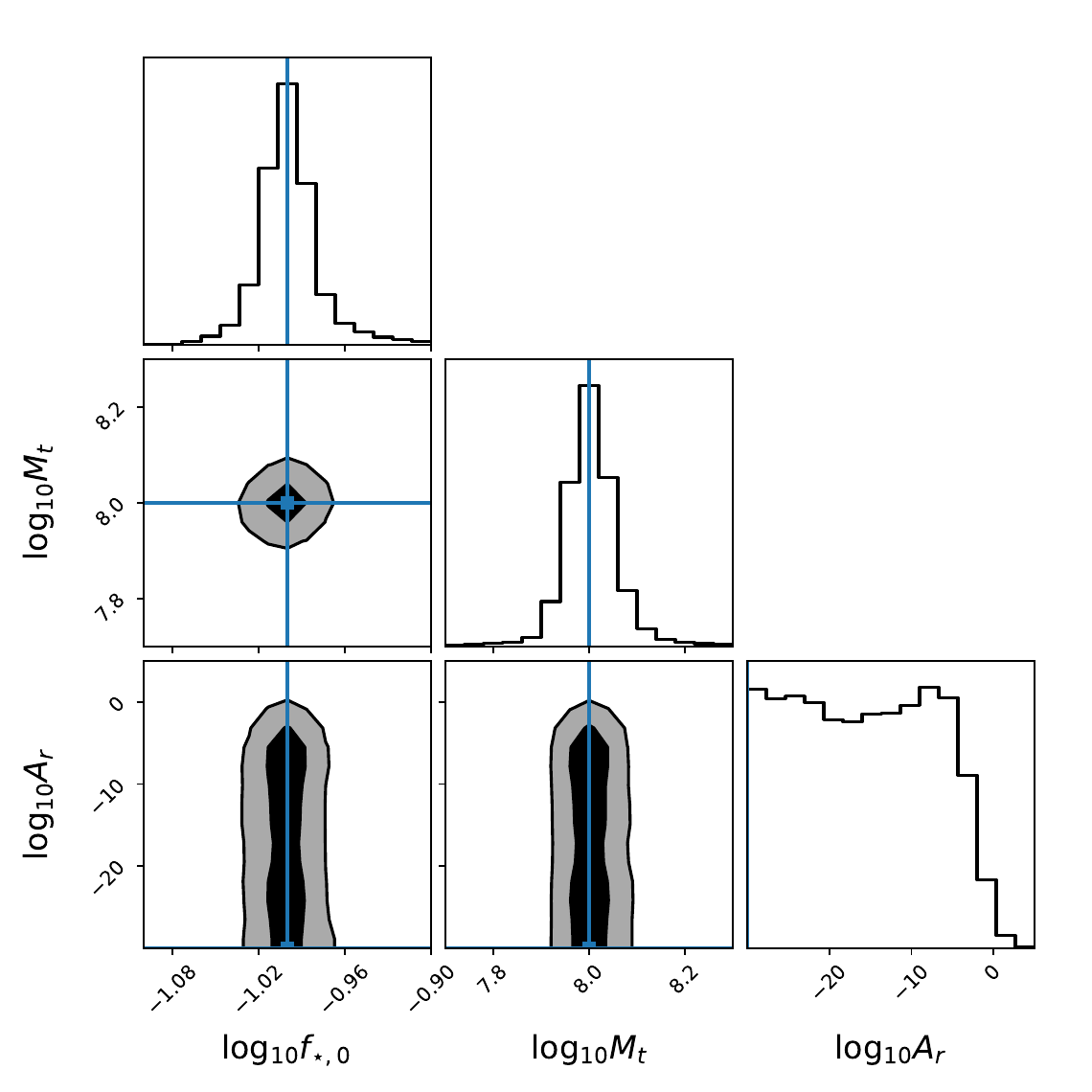}
\caption{Posterior distribution on the model parameters expected from upcoming SKA-Low measurements with 1000 hour observing time. The assumed parameter values for the mock observation is shown with blue lines. The dark and light contours represent the 68\% and 95\% confidence level. We see that this measurement will not only constrain the astrophysical processes, it will provide strong upper limits on the excess radio background (log$_\mathrm{10}\Ar$) that can be translated into constraints on the HFGWs.
}
\label{fig:corner_mcmc}
\end{figure}

To forecast the constraining capability of SKA, we use a Bayesian framework that provides the posterior  distribution $\mathcal{P}(\pmb{\theta}|\mathrm{d})$ quantifying the probability of model parameters $\pmb{\theta}$ given data vector $\pmb{d}$. This quantity is given as follows,
\begin{equation}
    \mathcal{P}(\pmb{\theta}|\mathrm{d}) \propto \mathcal{L}(\pmb{d}|\pmb{\theta}) \pi(\pmb{\theta}) \ ,
\end{equation}
where $\mathcal{L}(\pmb{d}|\pmb{\theta})$ and $\pi(\pmb{\theta})$ are the likelihood and prior distribution respectively. We assume a Gaussian likelihood that is given in equation~(21) of Ref.~\cite{giri2022imprints}. 
This likelihood incorporates the impact of both the cosmic variance and system noise expected from 1000 hour with SKA observation.

We use an MCMC sampler implemented in the publicly available \texttt{emcee} package\footnote{The package can be found at \url{https://emcee.readthedocs.io/en/stable/}.}~\cite{foreman2013emcee} for exploring  model parameter space. 
Below we describe these model parameter along with the assumed prior range for each of them.
\begin{enumerate}
    \item $f_{*,0}$: The amplitude of the star formation rate in the dark matter haloes hosting photon sources during these early times. The number of photons emitted is proportional to the stellar mass. Therefore we can control the amount of emitted photons with this parameter. We consider a flat prior between 0.01 and 1 in log-space, 
    which is consistent with models required to interpret the ultraviolet luminosity function measurements \cite[e.g.][]{sun2023bursty,dayal2024warm}
    . We assume $f_{*,0}=0.1$ for producing the mock observation. 
    \item $M_{t}$: The minimum dark matter mass that can sustain star formation. We consider a flat prior between $3.2\times 10^6$ and $3.2\times 10^9~M_\odot/h$ in log-space. We assume $M_{t}=10^{8}~M_\odot/h$, which is close to the threshold that can sustain source formation due to molecular cooling~\cite{tegmark1997small,nebrin2023starbursts}.
    \item $\Ar$: The parameter describing the magnitude of excess radio background corresponding to the 21-cm signal. In order to explore interesting values shown in Figure~\ref{fig:Ar_f_OmGW}, we consider a flat prior between $-30$ and $30$ in log-space. We produced the mock observation assuming absence of excess radio background that is quantified as $\Ar=10^{-30}$.
\end{enumerate}
As the goal of this study is to study the constraints in the HFGWs that depends only on $A_r$, we have chosen a simple two parameter astrophysical model to describe the photon sources during cosmic dawn.
In Figure~\ref{fig:corner_mcmc}, we show the full corner plot\footnote{The plot is produced using the \texttt{corner} package~\cite{corner} that can be found at \url{https://corner.readthedocs.io/en/latest/}.} for the posterior distribution from the MCMC run. We find that the ground truth for the two astrophysical parameters are predicted correctly at both the 68\% and 95\% confidence level. We estimated an upper limit on the excess radio background with $\Ar\lesssim 10^{-9}$ at 68\% confidence level from the 1D marginalised posterior distribution. 

\bibliographystyle{JCAP}
\bibliography{references}

\end{document}